\documentclass[%
 reprint,
superscriptaddress,
 amsmath,amssymb,
 aps,
]{revtex4-2}

\usepackage{graphicx}
\usepackage{dcolumn}
\usepackage{bm}
\usepackage{hyperref}
\hypersetup{colorlinks,linkcolor={blue},citecolor={blue},urlcolor={black}}

\begin{document}

\preprint{APS/123-QED}

\title{Empowering Control of Antiferromagnets by THz-induced Spin Coherence }
\author{T.G.H. Blank}
\affiliation{Radboud University, Institute for Molecules and Materials, 6525 AJ Nijmegen, the Netherlands.}
\author{K.A. Grishunin}
\affiliation{Radboud University, Institute for Molecules and Materials, 6525 AJ Nijmegen, the Netherlands.}
\author{B.A. Ivanov}
\affiliation{Radboud University, Institute for Molecules and Materials, 6525 AJ Nijmegen, the Netherlands.}
\affiliation{Institute of Magnetism, National Academy of Sciences and Ministry of Education and Science, Kiev, Ukraine.}
\author{E.A. Mashkovich}
\affiliation{University of Cologne, Institute of Physics II, Cologne D-50937, Germany.}
\author{D. Afanasiev}
\affiliation{Radboud University, Institute for Molecules and Materials, 6525 AJ Nijmegen, the Netherlands.}
\author{A.V. Kimel}
\affiliation{Radboud University, Institute for Molecules and Materials, 6525 AJ Nijmegen, the Netherlands.}

\date{\today}
\begin{abstract}
Finding efficient and ultrafast ways to control antiferromagnets is believed to be instrumental in unlocking their potential for magnetic devices operating at THz frequencies. Still, it is challenged by the absence of net magnetization in the ground state. Here, we show that the magnetization emerging from a state of coherent spin precession in antiferromagnetic iron borate FeBO$_3$ can be used to enable the nonlinear coupling of light to another, otherwise weakly susceptible, mode of spin precession. This nonlinear mechanism can facilitate conceptually new ways of controlling antiferromagnetism.
\end{abstract}

\maketitle

Thermodynamic theory models a simple antiferromagnet as two ferromagnets with two mutually equal, but oppositely oriented magnetizations $\mathbf{M}_1$ and $\mathbf{M}_2$. As the net magnetization $\mathbf{M} = \mathbf{M}_1 + \mathbf{M}_2 $ is zero, the spin order is described by the nonzero antiferromagnetic N\'{e}el vector $\mathbf{L} = \mathbf{M}_1 - \mathbf{M}_2$. However, the N\'{e}el vector $\mathbf{L}$ in thermodynamic equilibrium is notoriously insusceptible to external magnetic fields~\cite{Nemec2018}. Despite the $60$-year-long search for thermodynamic field conjugates to the antiferromagnetic order parameter~\cite{Song_2018}, through which it may be altered, efficient mechanisms to control antiferromagnets are still at the focus of fundamental research, hampering further developments of antiferromagnetic spintronics, magnonics, and data storage~\cite{Jungwirth2016, RevModPhys.90.015005, Nemec2018}.

\begin{figure}[b!]
    \centering
    \includegraphics{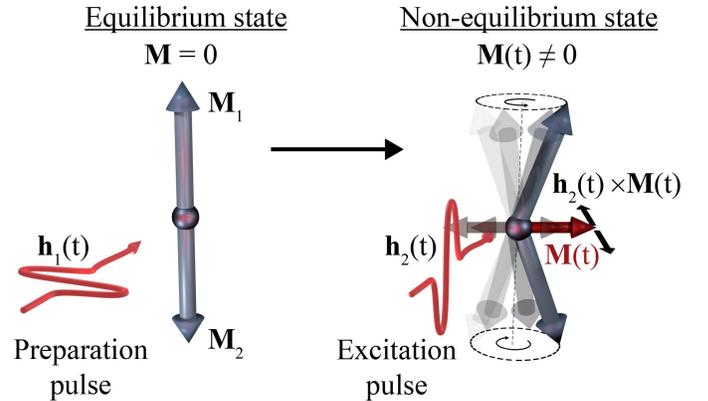}
    \caption{Enhanced susceptibility of antiferromagnets via a coherent magnonic state. (left) Magnetic configuration of a collinear antiferromagnet in the unperturbed state. The two arrows indicate the magnetizations of the antiferromagnetically coupled sublattices $\mathbf{M}_1$ and $\mathbf{M}_2$ with $\mathbf{M}_1 = - \mathbf{M}_2$ such that the net magnetization is absent $\mathbf{M} = 0$. However, by applying a ``preparation pulse'' with the magnetic field $\mathbf{h}_1(t)$ perpendicular to the spins, it is possible to generate a coherent magnonic state, inducing a dynamic magnetization $\mathbf{M}(t)$ as depicted by the spin arrows on the right. (right) Once in the non-equilibrium state, a second THz “excitation pulse” polarized along the equilibrium spin direction will now be able to exert a nonzero dynamic torque $\mathbf{h}_2(t) \times \mathbf{M}(t)$. The preparation pulse has thereby enhanced the susceptibility of the excitation pulse via a coherent magnonic state. }
    \label{fig:1}
\end{figure}

Out of equilibrium, the situation changes dramatically. Absolutely every antiferromagnet can be driven in a non-equilibrium coherent magnonic state with nonzero dynamic net magnetization $\mathbf{M}(t)$, by resonantly driving $\mathbf{L}(t)$ with a THz magnetic field $\mathbf{h}(t)$ applied perpendicular to the spins~\cite{Kampfrath2011}. The emerging dynamic net magnetization follows from elementary Lagrangian mechanics, which shows that \cite{Zvezdin1979, andreev1980symmetry, doi:10.1063/1.5041427}:
\begin{equation}
    \mathbf{M}(t) = \frac{1}{2\gamma H_{\mathrm{ex}}M_0} \left[\mathbf{L}\times\frac{d\mathbf{L}}{dt}\right],
\end{equation}
where $\gamma$ the gyromagnetic ratio, $H_{\mathrm{ex}}$ the antiferromagnetic exchange field and $M_0 \equiv |\mathbf{M}_{1,2}|$. Once in the coherent magnonic state, the antiferromagnet is characterized by an enhanced susceptibility to magnetic field, as the latter can couple to the dynamic magnetization $\mathbf{M}(t)$. Hence, when preparing an antiferromagnet in a coherent magnonic state first with a preparation pulse  $\mathbf{h}_1(t)$ applied perpendicular to the spins, the resulting dynamic magnetization greatly enhances the susceptibility to the magnetic field of a subsequent excitation pulse $\mathbf{h}_2(t)$ applied along the equilibrium spin direction (see Fig.~\ref{fig:1} for an illustration of this idea). This means that the response of an antiferromagnet to a pair of pulses of magnetic fields, $\mathbf{h}_1(t)$ and $\mathbf{h}_2(t)$, can be much larger than the effects caused by each of the pulses taken alone, and is therefore beyond trivial superposition.

\begin{figure}[t!]
    \centering
    \includegraphics{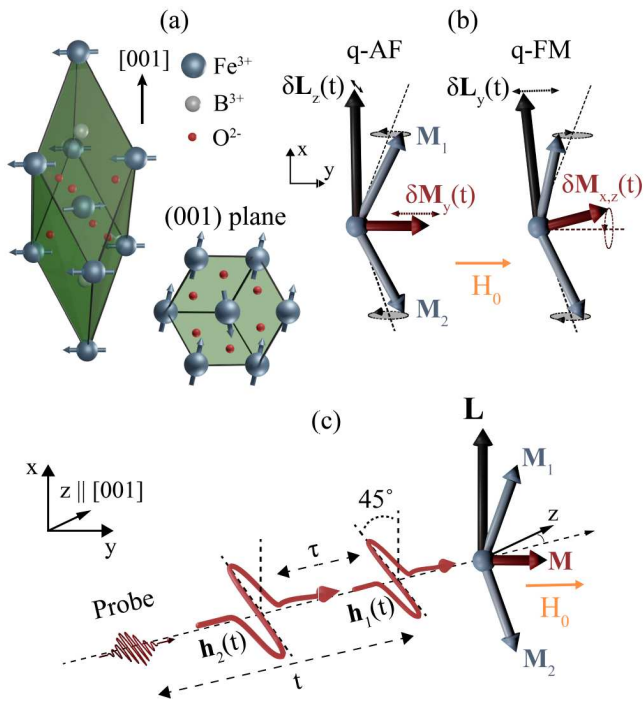}
    \caption{(a) Primitive unit cell of the FeBO$_3$ crystal with the Fe$^{3+}$ spins indicated by arrows. The spins, which lie in the $(001)$ easy plane of anisotropy, can be divided into two antiferromagnetically coupled sublattices $\mathbf{M}_1$ and $\mathbf{M}_2$ that are canted by DMI, resulting in a small net magnetization $\mathbf{M}$. (b) The two eigenmodes of spin precession of a canted antiferromagnet. The high-frequency q-AF mode results in longitudinal modulations of the net magnetization and can be excited by an oscillating THz magnetic field perpendicular to $\mathbf{L}$. The other mode, the low-frequency q-FM mode, corresponds to transverse precession of $\mathbf{M}$. (c) A schematic illustration of $2$D THz spectroscopy in FeBO$_3$. Both THz magnetic field pulses are polarized at $45^\circ$ from the net magnetization $\mathbf{M}$ to ensure there is a THz magnetic field component both orthogonal as well as parallel to the spins, as is required for the idea illustrated in Fig.~\ref{fig:1}. The probe pulse (pulse duration $\sim 100$~fs) electric field (central wavelength $800$~nm) was initially polarized along the $y$-axis. The dynamic magneto-optical polarization rotation of the probe was detected in a balanced detection scheme.}
    \label{fig:2}
\end{figure}

To explore this mechanism for empowering THz control of antiferromagnetism, we employ the principles of two-dimensional (2D) THz spectroscopy \cite{PhysRevLett.107.067401, Woerner_2013, PhysRevLett.118.207204, Lu2018, PhysRevLett.122.073901, mashkovich2021terahertz, blank2022two}. This spectroscopic technique involves a pair of intense THz pump pulses applied successively to the material, mutually separated by an \textit{excitation time} delay $\tau$. Using the pump-probe technique and measuring the rotation of polarization by magneto-optical effects of an ultrashort near-infrared probe pulse delayed by the \textit{detection time} $t$, one can trace magnetization dynamics $\mathbf{M}(t)$ in the form of probe polarization rotation $\theta(t)$ induced by both the combined action of the pump pulses $\theta_{12}(t,\tau)$, and by each of the pump pulses separately - $\theta_{1}(t,\tau)$  and $\theta_{2}(t,\tau)$. The nonlinear part of spin dynamics that stands beyond the trivial superposition can be readily extracted from the total signal by finding the difference
\begin{equation}
    \theta_{\mathrm{NL}}(t,\tau) = \theta_{12}(t,\tau) - \theta_{1}(t,\tau) - \theta_{2}(t,\tau).
\end{equation}

\begin{figure*}[t!]
    \centering
    \includegraphics{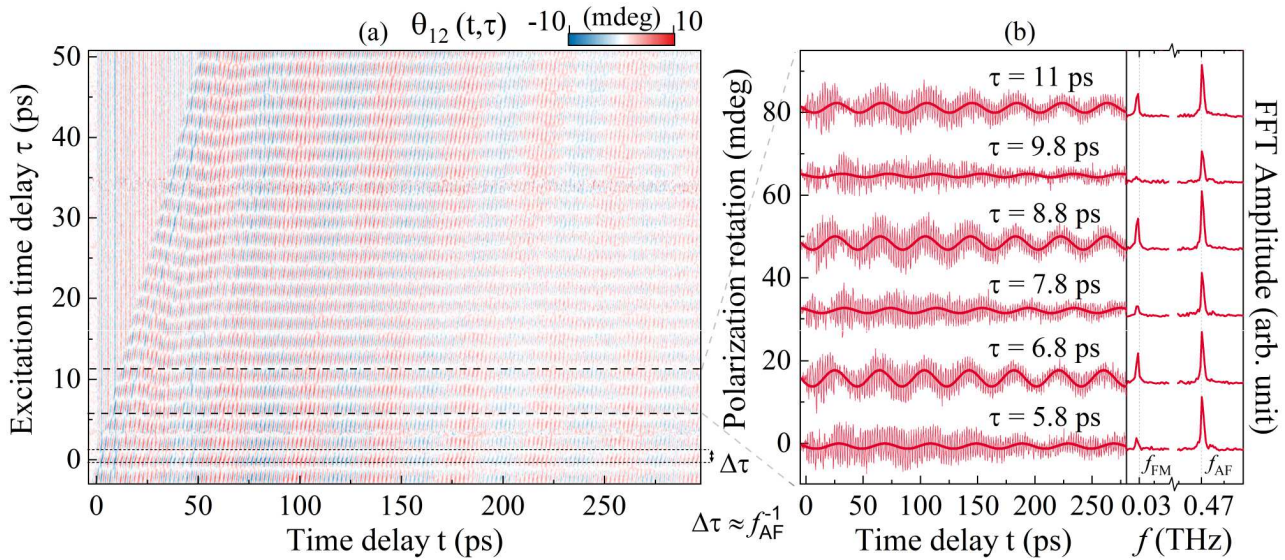}
    \caption{Time-domain result of $2$D THz spectroscopy in FeBO$_3$. (a) Double THz pump induced polarization rotation of the probe $\theta_{12}(t,\tau)$ as a function of the detection time delay $t$ and the excitation time delay $\tau$ (see Fig.~\ref{fig:2}(c)). The regions where the q-FM mode is seen to be quenched are separated by a characteristic time $\Delta \tau \approx f_{\mathrm{AF}}^{-1}$. (b) Cross-section of the two-dimensional graph and the associated FFT depicted for several $\tau$, emphasizing the toggling of the q-FM mode with a repetition rate that can be associated with the q-AF mode.}
    \label{fig:3}
\end{figure*}

Here we apply this technique to FeBO$_3$ - a prototypical antiferromagnet ideally suitable for time-resolved magneto-optical pump-probe experiments. For temperatures below the N\'{e}el point $T_\mathrm{N} \approx 348$~K, the Fe$^{3+}$ spins in this material form two equivalent antiferromagnetically coupled macroscopic sublattice magnetizations $\mathbf{M}_1$ and $\mathbf{M}_2$ lying in the $(001)$ sample plane \cite{doi:10.1063/1.1652682} (see Fig.~\ref{fig:2}(a)). In equilibrium the magnetization vectors are slightly canted at an angle of $\sim 1^\circ$ because of the Dzyaloshinskii–Moriya interaction (DMI) \cite{Dmitrienko2014}, resulting in a small net magnetization $\mathbf{M}$ perpendicular to $\mathbf{L}$. The material hosts two non-degenerate mutually orthogonal modes of spin precession, with significantly different frequencies: the quasi-antiferromagnetic (q-AF) and quasi-ferromagnetic (q-FM) modes \cite{velikov1974antiferromagnetic} (see Fig.~\ref{fig:2}(b)). The q-AF mode essentially results in out-of-plane motion of the N\'{e}el vector and in longitudinal dynamics of the net magnetization $\mathbf{M}$ ($f_\mathrm{AF} \approx 0.5$~THz at $T = 78$~K), while the q-FM mode ($f_\mathrm{FM} \approx 0.03$~THz at $T = 78$~K) is mainly associated with in-plane dynamics of $\mathbf{L}$ and results in transverse precessional dynamics of $\mathbf{M}$ \cite{1059040}. Most importantly, the material is characterized by strong magneto-optical effects which facilitate sensitive detection of magnetization dynamics $\mathbf{M}$(t) with the help of light \cite{doi:10.1063/1.1652682, doi:10.1063/1.1658881, PhysRevLett.89.287401, PhysRevLett.99.167205}. Finally, the choice of FeBO$_3$ is motivated by recent experiments which show that an intense single-cycle THz pulse can excite both modes simultaneously \cite{PhysRevLett.123.157202}. Here, the amplitude of the q-AF mode scaled linearly while that of the q-FM scaled quadratically with the THz magnetic field. The linear excitation of the q-AF mode could be explained in terms of a magnetic-dipole interaction or Zeeman torque of the THz magnetic field with spins~\cite{PhysRevB.104.024419}, the mechanism of the nonlinear excitation of the q-FM mode is still unknown. 

The THz pulses used for $2$D spectroscopy were generated by titled pulse front optical rectification in a LiNbO$_3$ crystal \cite{Hebling:02, doi:10.1063/1.2734374, doi:10.1063/1.3560062}, using two beams of amplified $100$~fs laser pulses at a central wavelength of $800$~nm. The THz waveforms were mapped and calibrated by electro-optical sampling in GaP (see Supplemental Material~\footnote{\label{footnote}See the Supplemental Material for details on the calibration of THz pulses in GaP, details of $2$D spectroscopy, the full analytical treatment of the sigma-model, the explanation of the Doppler-like shift seen in the $2$D~FFT diagrams and for all supplemental experimental and numerical results, which includes Refs.~\cite{Hebling:02, doi:10.1063/1.3560062, mashkovich2021terahertz, doi:10.1063/1.5041427, doi:10.1063/1.4865565,  PhysRevLett.123.157202, RUDASHEVSKY1972959, Kimel2009, Galkin2008, LandauLifshitz}.}), resulting in peak electric fields of $400$~kV/cm corresponding to about $130$~mT peak magnetic field. The two THz pulses were incident on the sample approximately along the [$001$] crystallographic axis, but at a slight angle ($\sim 12^\circ$) away from this axis by tilting the sample as in Ref.~\cite{PhysRevLett.123.157202}. The FeBO$_3$ sample was cooled down to $78$~K where the observed dynamics are optimal. An external magnetic field $\mu_0 \mathbf{H}_0 \approx 70$~mT was applied predominantly in the $(001)$ easy plane to control the in-plane orientation of $\mathbf{M}$. The direction of the aligned magnetization is referred to as the experimental $y$-axis. The experimental scenario is illustrated in Fig.~\ref{fig:2}(c). The polarization of the THz magnetic fields could be controlled by a set of wire-grid polarizers and was set to $45^\circ$ with respect to the $y$-axis. Note that this rotation using wire-grid polarizers reduces the peak THz magnetic field to about $110$~mT. By setting this particular THz polarization, we ensured that the first THz pulse contains a strong THz magnetic field component perpendicular to the spins to excite the q-AF mode \cite{PhysRevB.104.024419}, while the second pulse has a large component perpendicular to the anticipated emergent dynamic magnetization as illustrated in Fig.~\ref{fig:1}.

Figure~\ref{fig:3} summarizes the results of the $2$D time-resolved measurements. In particular, Fig.~\ref{fig:3}(a) shows the calibrated magnetic response from the combined action of the two THz pulses $\theta_{12}(t,\tau)$. It is seen that fast oscillations in time $t$, which can be assigned to the q-AF mode, are intertwined with slower oscillations at the frequency typical for the q-FM mode $f_{\mathrm{FM}}$. A striking feature of this data is the periodic dependence of the amplitude of the low frequency (q-FM) mode as a function of the excitation time delay $\tau$, which can be seen in the $2$D mapping as equidistant white horizontal stripes. According to the color code, the white color corresponds to zero amplitude. This is seen even more directly in Fig.~\ref{fig:3}(b), which shows a selection of time traces obtained at different $\tau$. It is seen that subtle changes in the excitation time delay $\tau \ll f_{\mathrm{FM}}^{-1}$ have a dramatic impact on the amplitude of the mode, practically switching it ``on'' and ``off''. In fact, the toggling of the q-FM mode as a function of $\tau$ has a periodicity $\Delta \tau$ that can be related to the frequency of the q-AF mode $\Delta \tau \approx f_{\mathrm{AF}}^{-1}$, matching the mechanism illustrated in Fig.~\ref{fig:1}. 

\begin{figure*}[t!]
    \centering
    \includegraphics{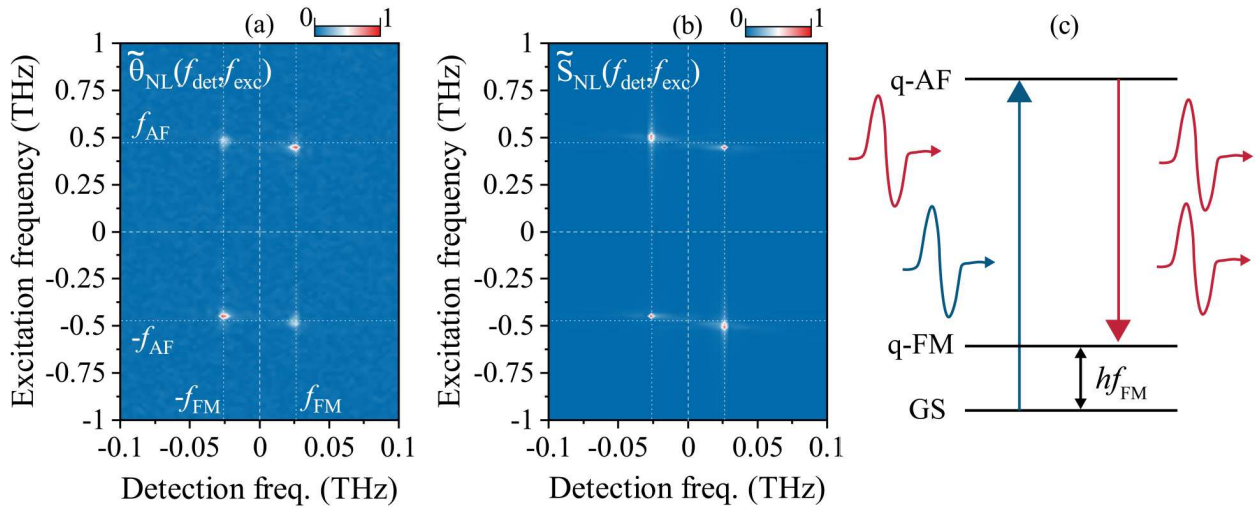}
    \caption{(a) (left panel) $2$D~FFT of the nonlinear part of the experimental data. The off-diagonal peaks indicate a nonlinear transfer of energy from the q-AF mode to the q-FM mode, supporting the result of Fig.~\ref{fig:3}. (b) The $2$D~FFT of the non-linear part of the simulated data $S(t,\tau) = l_y(t,\tau) + l_z(t,\tau)$, which has a large qualitative agreement with the experiment. (c) Energy diagram of magnonic stimulated Raman scattering. Here, the first THz pulse excites the q-AF magnon and the second THz pulse stimulates the Stokes transition to the q-FM magnon state. Analogously, the nonlinear excitation of the q-FM mode with a single THz pulse can be regarded as magnonic impulsive stimulated Raman scattering. }
    \label{fig:4}
\end{figure*}

The $2$D time-resolved data of Fig.~\ref{fig:3}(a) can also be presented in reciprocal space using the two-dimensional fast Fourier transform ($2$D~FFT). The conjugate frequencies to the times $t$ and $\tau$ are referred to as \textit{detection frequency} $f_{\mathrm{det}}$ and \textit{excitation frequency} $f_{\mathrm{ex}}$, respectively. Figure~\ref{fig:4}(a) shows the $2$D Fourier amplitude for the non-linear signal $\tilde{\theta}_{\mathrm{NL}}(f_{\mathrm{det}}, f_{\mathrm{ex}})$, which clearly reveals maxima at $(f_{\mathrm{det}}, f_{\mathrm{ex}}) \approx (\pm f_{\mathrm{FM}}, \pm f_{\mathrm{AF}} )$. Note that due to peculiarities of $2$D THz spectroscopy and the corresponding $2$D~FFT, the peaks are slightly shifted in the excitation frequency $|f_{\mathrm{ex}}| = |f_{\mathrm{AF}} \pm f_{\mathrm{FM}}|$. As explained in detail in the Supplemental Material~\cite{Note1}, this shift can be attributed to an effect similar to the Doppler effect. In general, the result of Fig.~\ref{fig:4}(a) implies nonlinear energy transfer between the modes \cite{mashkovich2021terahertz} and clearly reveals that the nonlinear excitation of the low-frequency q-FM mode is mediated by the second and seemingly orthogonal q-AF mode.

To formally describe the discovered nonlinear coupling between q-FM and q-AF modes, we employ the Lagrangian formulation of the sigma model (see Refs.~\cite{doi:10.1063/1.5041427, andreev1980symmetry, Zvezdin1979} and Supplemental Material~\cite{Note1}). Usually, this Lagrangian $\mathcal{L}$ is only treated in linear approximation, in which case it can be divided into two mutually independent parts $\mathcal{L} \approx \mathcal{L}_{y,0}$ + $\mathcal{L}_{z,0}$,
\begin{equation}
    \begin{aligned}
    \label{eq}
    \frac{2\gamma H_{\mathrm{ex}}}{\hbar}\mathcal{L}_{y,0} = \left(\frac{\partial l_y}{\partial t}\right)^2 - \omega^2_{\mathrm{FM}} l_y^2 + 2\gamma^2 H_{\mathrm{eff}}l_yh_x, \\ 
    \frac{2\gamma H_{\mathrm{ex}}}{\hbar}\mathcal{L}_{z,0} = \left(\frac{\partial l_z}{\partial t}\right)^2 - \omega^2_{\mathrm{AF}} l_z^2 - 2\gamma l_z \frac{\mathrm{d}h_y}{\mathrm{d}t}.
    \end{aligned}
\end{equation}
Here $H_{\mathrm{eff}} = H_0 + H_\mathrm{D}$ with $H_\mathrm{D}$ the DMI field, $\hbar$ is Planck's constant and $\omega_{\mathrm{FM}}$ and $\omega_{\mathrm{AF}}$ are the angular frequencies of the magnetic modes. The first term $\mathcal{L}_{y,0}$ describes the dynamics of the  $y$-projection of normalized N\'{e}el vector $\mathbf{l} = \mathbf{L} / |\mathbf{L}|$ and thus corresponds to the q-FM mode, while $\mathcal{L}_{z,0}$ describes dynamics of the $z$-projection that can be assigned to the q-AF mode (see Fig.~\ref{fig:2}(b)). The equations of motion derived from these Lagrangians are not coupled (see Supplemental Material~\cite{Note1}) and possess a substantial difference regarding the THz excitation. The dynamics of the z-component $l_z(t)$, i.e. the q-AF mode, can be triggered efficiently by the THz magnetic field when this field is aligned perpendicular to the spins $\mathbf{h} \perp \mathbf{L}$~\cite{PhysRevB.104.024419}. The amplitude of the resulting oscillations in $l_z(t)$ will be a linear function of the THz magnetic field. This linear mechanism is, in principle, also able to excite the q-FM mode when the THz magnetic field is aligned perpendicular to the weak magnetization $\mathbf{M}$. However, such a direct excitation is expected to be rather inefficient due to the absence of a strong spectral component at the q-FM resonance frequency in the THz excitation spectrum (see Supplemental Material~\cite{Note1}). Therefore, despite being theoretically possible, no signature of such a linear excitation of the q-FM mode was observed experimentally either in this work or before \cite{PhysRevLett.123.157202}, showing that resonant excitation of the q-FM mode in our experiment can be essentially neglected.

It is thus obvious that in order to explain the excitation of the q-FM mode via coupling to the q-AF mode, we need to go beyond the linear approximation. To this end, we add to the Lagrangian the non-linear coupling term of the lowest order $\mathcal{L}_{\mathrm{coupl}}$:
\begin{equation}
    \frac{H_{\mathrm{ex}}}{\hbar}\mathcal{L}_{\mathrm{coupl}} = h_x\left(l_z\frac{\mathrm{d}l_y}{\mathrm{d}t} - l_y \frac{\mathrm{d}l_z}{\mathrm{d}t}\right).
\end{equation}
This term allows for coupling between the modes and the THz magnetic field. To show that this term is, in fact, responsible for the observed excitation of the q-FM mode, we derived the corresponding equations of motion for $l_x(t)$ and $l_y(t)$ and numerically solved the magnetization dynamics triggered by two pulses of THz magnetic field $\mathbf{h}(t,\tau) = \mathbf{h}_1(t) + \mathbf{h}_2(t,\tau)$ with modeled waveforms similar to the experimental waveforms (see Supplemental Material~\cite{Note1} for more details). Figure~\ref{fig:4}(b) shows the 2D~FFT of the nonlinear part of the simulated signal $S(t,\tau) = l_y(t,\tau) + l_z(t,\tau)$. The result of the simulations has a good qualitative agreement with the experiment and allows us to conclude that this nonlinear term adequately describes the observed coupling between otherwise non-interacting modes of spin precession in FeBO$_3$.

Using this theory, we could analytically estimate how much the susceptibility of the q-FM mode to a THz magnetic field pulse can actually be enhanced by a preparation pulse (see Fig.~\ref{fig:1}). We expressed this enhancement by the amplification factor $\mu$ defined as the ratio of excitation efficiency of the q-FM mode by the excitation pulse with and without the preparation pulse present (see Fig.~\ref{fig:1}). The analytical expression for $\mu$ has been given in the Supplemental Material~\cite{Note1}. In the case of FeBO$_3$, having a relatively large spin canting in the ground state, and for moderate THz magnetic fields $|\mathbf{h}_{1,2}|\sim 0.1$~T, the amplification factor is estimated to be nearly an order of magnitude $\mu \sim 7$. Simple estimates show that when applying higher but realistic THz magnetic fields $>0.3$~T \cite{doi:10.1063/1.3560062, PhysRevApplied.19.034018} to antiferromagnets with a smaller spin canting in the ground state, the enhancement can be boosted by multiple orders of magnitude. In the case of collinear antiferromagnets such as NiO, the enhancement of the susceptibility is by definition infinite as these materials have zero magnetization in the ground state. As the employed model is very general, one can boost the susceptibility of spins in absolutely any antiferromagnet, with and without spin canting, by a non-equilibrium coherent magnonic state.

To conclude, our results show that a coherent magnonic state can substantially change the properties of an antiferromagnet, enabling a new nonlinear path of controlling spins by a pair of THz pulses. This has been demonstrated by showing that a coherent q-AF magnonic state in FeBO$_3$ mediates the excitation of the q-FM mode by THz magnetic field. The effect is analogous to electronic or ionic Raman scattering \cite{Forst2011}, but involves exclusively magnonic excitations and can be thus called magnonic Raman scattering or THz-mediated magnon-magnon coupling, as illustrated in Fig.~\ref{fig:4}(c). Our work shows that although the efficient control of antiferromagnetism in thermodynamic equilibrium is still a challenge, the problem can be solved by pushing antiferromagnets into a non-equilibrium state where the susceptibility of spins to an external magnetic field is boosted. By combining various magnonic, phononic \cite{Disa2020, Afanasiev2021}, and electronic excitations \cite{nonspincontrol}, one can generate and explore diverse non-equilibrium states and find which excitation (or a combination thereof) facilitates the fastest and the most energy-efficient control of antiferromagnetism.

\begin{acknowledgments}
The authors thank S. Semin and C. Berkhout for their technical support. The work was supported by de Nederlandse Organisatie voor Wetenschappelijk Onderzoek (NWO) and the European Research Council ERC Grant Agreement No.101054664 273 (SPARTACUS). The contribution of E.A. Mashkovich has been funded by the Deutsche Forschungsgemeinschaft (DFG, German Research Foundation) - Project number 277146847 - CRC 1238.
\end{acknowledgments}


\bibliography{references}

\end{document}


\title{Supplemental Material for ``Empowering Control of Antiferromagnets by THz-induced Spin Coherence''}
\author{T.G.H. Blank}
\affiliation{Radboud University, Institute for Molecules and Materials, 6525 AJ Nijmegen, the Netherlands.}
\author{K.A. Grishunin}
\affiliation{Radboud University, Institute for Molecules and Materials, 6525 AJ Nijmegen, the Netherlands.}
\author{B.A. Ivanov}
\affiliation{Radboud University, Institute for Molecules and Materials, 6525 AJ Nijmegen, the Netherlands.}
\affiliation{Institute of Magnetism, National Academy of Sciences and Ministry of Education and Science, Kiev, Ukraine.}
\author{E.A. Mashkovich}
\affiliation{University of Cologne, Institute of Physics II, Cologne D-50937, Germany.}
\author{D. Afanasiev}
\affiliation{Radboud University, Institute for Molecules and Materials, 6525 AJ Nijmegen, the Netherlands.}
\author{A.V. Kimel}
\affiliation{Radboud University, Institute for Molecules and Materials, 6525 AJ Nijmegen, the Netherlands.}

\date{\small{\today}}

\maketitle
\textbf{Table of contents}

\tableofcontents 

\newpage

\section{2D THz spectroscopy, THz pulse calibration and experimental conditions}\label{sec:2Dspectroscopy}
\begin{figure}[h!]
    \centering
    \includegraphics{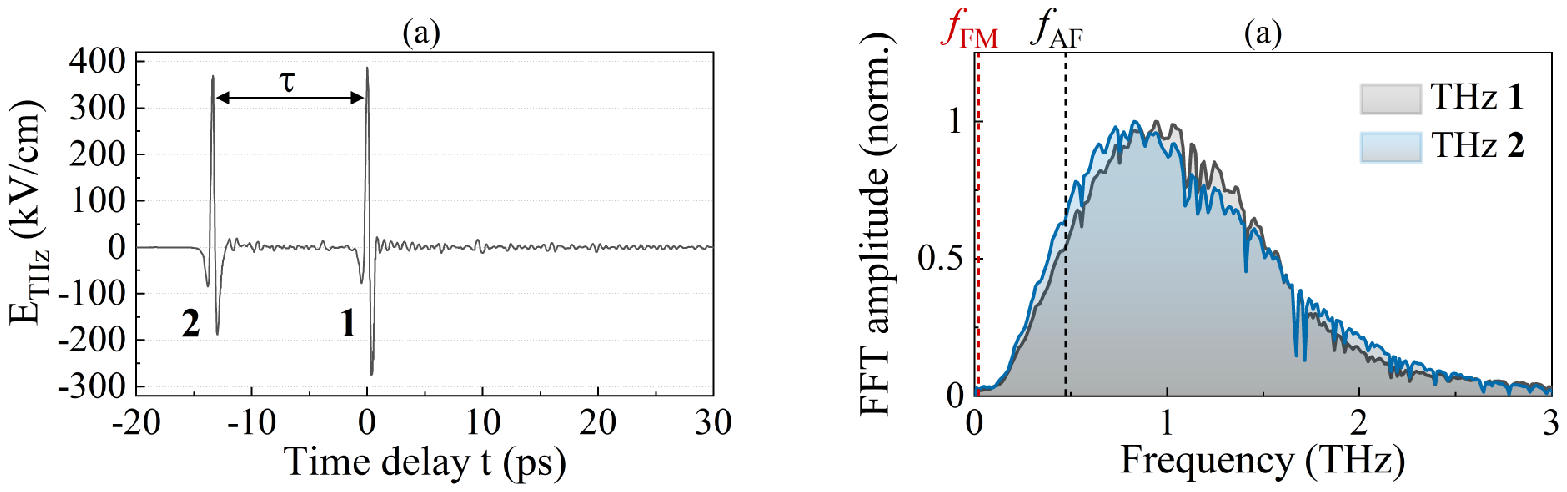}
    \caption{\textit{\small{(a) Time-resolved waveforms of the two THz pump pulses measured by electro-optical sampling in GaP. The two THz pulses, labeled $1$ and $2$, are separated by the excitation time delay $\tau$ and have peak electric fields of approximately $388$~kV/cm and $370$~kV/cm. After generation, the THz electric fields are initially polarized along the experimental $x$-axis (see Fig.~2 of the main article). In the experiment, we rotated the THz pulses by $45^\circ$ using two wire-grid polarizers, which as a consequence reduces the THz electric field by a factor of $0.854$. (b) Normalized FFT spectra of the two THz pulses. Although it can be seen in the time domain that the two waveforms are slightly dissimilar, the FFT spectra of the two THz pulses are nearly identical. The frequencies of q-FM and q-AF modes are marked in the excitation spectrum with red and black dotted lines. }}}
    \label{fig:2DTHz callib}
\end{figure}
The linearly polarized THz pulses were generated by tilted pulse front optical rectification in a LiNbO$_3$ crystal~\cite{Hebling:02, doi:10.1063/1.3560062}, using two beams of amplified $100$~fs laser pulses at a central wavelength of $800$~nm. The entire generation process was performed in nitrogen-purified air to avoid water absorption of the THz light. The relative delay or excitation time delay $\tau$ between the two THz pulses was controlled by a mechanical delay stage. Empirically, it was found that the superposition of the generation process does not hold for excitation time delays below $|\tau|<1$~ps. Here, the simultaneous presence of two laser pulses in the generation crystal causes mutual influence on the THz generation, and we excluded this part for the FFT analysis in Fig.~$4$ of the main article to avoid artificial signals. The generated pulses first move through a pair of wire-grid polarizers to control their power and polarization and were then expanded, collimated, and focused onto the sample using three off-axis parabolic mirrors. Here, the pulses overlap both in space and in time with a tightly focused (beam radius $50$~$\mu$m) weak ($\sim 1.5 \cdot 10^{-4}$~$\mu$J) linear polarized (electric field along experimental $y$-axis) probe pulse at the central wavelength of $800$~nm, whose time delay $t$ with respect to the first THz pulse was controlled using another mechanical delay stage. The THz-induced polarization changes of the probe were detected using a balanced detection scheme. The repetition rate of the probe pulses was $1$~kHz, and these of the two THz pulses (labeled ``$1$'' and ``$2$'') were controlled at $500$ and $250$~Hz, respectively, by mechanical choppers. This measurement scheme allowed us to extract the induced polarization changes of the probe by the two THz pulses $\theta_1(t,\tau)$ and $\theta_2(t,\tau)$ separately, as well as their combined signal $\theta_{12}(t,\tau)$, using a data-acquisition card as is described in Ref.~\cite{mashkovich2021terahertz}. By replacing the sample with a $50$~$\mu$m thick [$110$]-oriented GaP crystal, we mapped the calibrated THz waveforms by electro-optical sampling which yielded almost $400$~kV/cm peak electric field (corresponding to about $130$~mT magnetic field) for the two THz pulses as can be seen in Fig.~\ref{fig:2DTHz callib}. Rotating the THz polarization $45^\circ$ with the help of wire-grid polarizers reduces the THz amplitude by a factor $0.854$, hence the peak magnetic field in the experiment is approximately $110$~mT. 

As was mentioned in the main article, it is clearly visible in the FFT spectrum (Fig.~\ref{fig:2DTHz callib}(b)) that the spectral component of the THz field at the q-FM frequency is nearly absent, and much weaker than that at the q-AF frequency $\tilde{h}(f_{\mathrm{FM}}) \ll \tilde{h}(f_{\mathrm{AF}}) $.

\textit{Experimental conditions of the sample.} The sample normal coincided with the $(001)$ crystallographic axis and was tilted $\sim 12^\circ$ with respect to the optical axis of the collinearly traveling pump and probe pulses, to ensure that there is a projection of the dynamic magnetization $\mathbf{M}$ for both modes of spin resonance along the optical axis for the purpose of detection through magneto-optics~\cite{PhysRevLett.123.157202}. Therefore, the pump and probe do not travel exactly along the ($001$) axis, but are traveling at an angle of $\sim 12^\circ$ from this ($z$-)axis. This also means that the THz fields have a small projection out of the ($001$) plane, which is included in the analysis of next Section (see $h_z(t)$ in Eq.~\eqref{eq:9}). An external magnetic field $\mu_0 \mathbf{H}_0$ of $70$~mT was applied perpendicular to the optical axes, such that also this field was not exactly in the sample plane, but was tilted in the $yz$~plane by the same angle of $\sim 12^\circ$. However, given that the field is weak, the magnetization is also weak, and the angle is small, the out-of-plane projection of the magnetic field is negligible and only the in-plane $y$-projection of the field, which aligns the magnetization along the experimental $y$-axis, matters. The sample was mounted on a cold finger nitrogen flow cryostat and was kept at a constant temperature of $78$~K.

\section{Doppler shift in the 2D~FFT}\label{sec:geometry}
The appearance of a peak in the $2$D FFT spectrum (Fig.~4(a) of the main article) in either quadrant is determined by the direction of the wavefront in the time domain, because, contrary to the one-dimensional case, Fourier components in $2$D do not only have amplitude and phase but also direction. We are only required to examine the first (I) and fourth (IV) quadrants, as the data is purely real which implies that  the Fourier amplitude has inversion symmetry $|\tilde{\theta}_{\mathrm{NL}}(f_{\mathrm{det}}, f_{\mathrm{ex}})| = |\tilde{\theta}_{\mathrm{NL}}(-f_{\mathrm{det}}, -f_{\mathrm{ex}})|$. When the periodic wavefront travels in the top-right direction it appears in the first quadrant, while it appears in the fourth (IV) quadrant for bottom-right traveling waves.

\begin{figure}[h!]
    \centering
    \includegraphics[width = 0.7\textwidth]{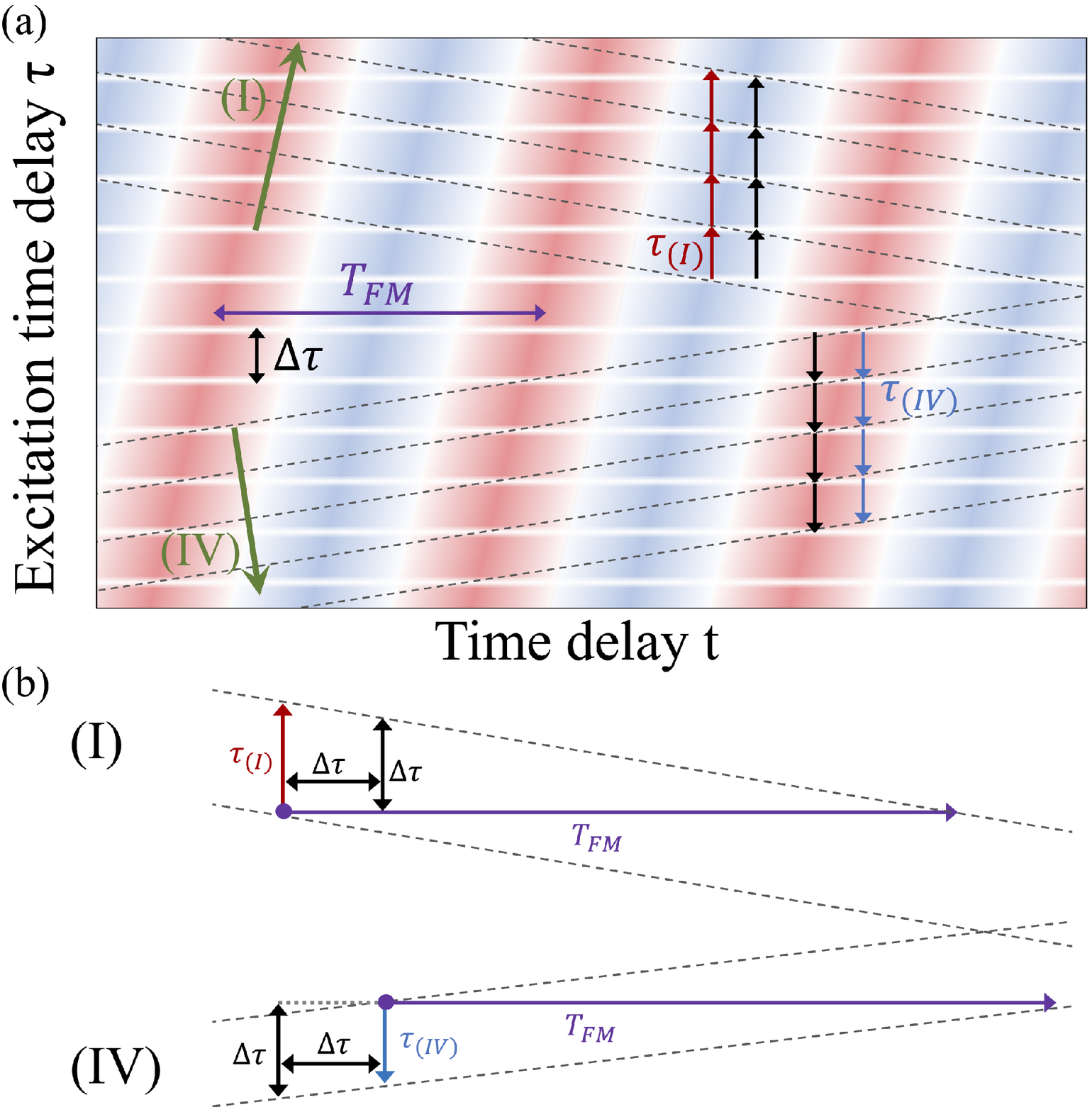}
    \caption{\textit{Interpretation of the frequency shifts observed in the $2$D~FFT diagrams. (a) Schematic of the time-domain data, showing the slow q-FM modulations together with the white lines of quenching, that are separated by a time $\Delta \tau = f_{\mathrm{AF}}^{-1}$. The wavefronts associated with the nonlinear peaks in the $2$D~FFT diagram are indicated by the dotted lines, and their propagation direction determines in which quadrant they appear (I or IV). From here, it can be seen why the peaks are slightly shifted away from $f_{\mathrm{AF}}$. The shift can be calculated by using the pictures in (b).}}
    \label{fig:geometry}
\end{figure}

It is possible to visualize how the two peaks $(f_{\mathrm{det}}, f_{\mathrm{ex}}) \approx (f_{\mathrm{FM}}, \pm f_{\mathrm{AF}})$ arise, by drawing the wavefronts of periodicity as was done in the schematic illustration of the experimental data in Fig.~\ref{fig:geometry}(a), where $T_{\mathrm{FM}} = f^{-1}_{\mathrm{FM}}$ the period of the q-FM mode and $\Delta \tau$ is the spacing between the white lines which was experimentally determined to be related to the q-AF mode frequency $\Delta \tau = f_{\mathrm{AF}}^{-1}$. From here, it can be seen why the peak in the first quadrant is slightly red-shifted from the q-AF resonance frequency, as the periodicity in this quadrant is $\tau_{\mathrm{(I)}} > \Delta \tau = f_{\mathrm{AF}}^{-1}$. Similarly, it can be seen why the peak in the fourth quadrant is blue-shifted $\tau_{\mathrm{(IV)}} < \Delta \tau = f_{\mathrm{AF}}^{-1}$. The exact shifts can be calculated by noting that the starting point of the q-FM mode is set by the arrival of the second THz pulse, which excites the mode from the non-equilibrium state that was triggered by the first THz pulse. In other words, the slope of the blue and red regions in Fig.~\ref{fig:geometry}(a) are diagonal, even although it might not appear as such in Fig.~\ref{fig:geometry}(a) or Fig.~\ref{fig:Exp_TD} given that the scales on the two axes are unequal. This fact gives
\begin{equation}
    \tau_{\mathrm{(I)}} = \frac{\Delta \tau T_{\mathrm{FM}}}{T_{\mathrm{FM}} - \Delta \tau}, \quad \quad \quad \tau_{\mathrm{(IV)}} = \frac{\Delta \tau T_{\mathrm{FM}}}{T_{\mathrm{FM}} + \Delta \tau},
\end{equation}
as was calculated using the geometric picture of Fig.~\ref{fig:geometry}(b). From here, we can find the shifted excitation frequencies of the two peaks in the first and fourth quadrants:
\begin{equation}
    f_{\mathrm{(I)}} \equiv  \tau_{\mathrm{(I)}}^{-1} = f_{\mathrm{AF}} - f_{\mathrm{FM}}, \quad \quad \quad f_{\mathrm{(IV)}} \equiv  \tau_{\mathrm{(IV)}}^{-1} = f_{\mathrm{AF}} + f_{\mathrm{FM}}.
\end{equation}
The same prediction can be made by the analytical solution of the sigma model, which will be treated in the next Section. Experimentally, one could get rid of the shift by defining an inertial frame where the detection time $t$ is defined from the point of overlap of the two THz pulses (instead of defining $t$ with respect to the first THz pulse). The two THz pulses must then be delayed from this fixed point by $\pm \frac{\tau}{2}$ such that their mutual delay is again the excitation time delay $\tau$.

\newpage

\section{Analytical sigma model}
For a two-sublattice antiferromagnet with sublattice magnetizations $\mathbf{M}_1$ and $\mathbf{M}_2$, we define the unit N\'{e}el vector $\mathbf{l} \equiv (\mathbf{M}_1 - \mathbf{M}_2 )/2M_0$ and the reduced magnetization $\mathbf{m}=( \mathbf{M}_1 + \mathbf{M}_2 )/2 M_0$, where $|\mathbf{M}_1| =|\mathbf{M}_2| = M_0$. By definition, these vectors are subject to the constraints $\mathbf{l}^2 +\mathbf{m}^2 = 1$ and $\mathbf{l}\cdot\mathbf{m} = 0$.

The magnetization dynamics for any antiferromagnet can be described by considering the dynamical equations of motion for the vector $\mathbf{l}$, known as the sigma model~\cite{doi:10.1063/1.5041427}. Within this model, $\mathbf{m}$ is a dependent variable of $\mathbf{l}$:
\begin{equation}\label{eq:3}
\mathbf{m} = \frac{1}{H_{\mathrm{ex}}}\left[\mathbf{H}_{\mathrm{eff}}-\mathbf{l}\left(\mathbf{H}_{\mathrm{eff}}\cdot\mathbf{l}\right)+\frac{1}{\gamma }\left(\frac{\partial\mathbf{l}}{\partial t}\times\mathbf{l}\right)\right] \\ 
\end{equation}
where the effective field $\mathbf{H}_{\mathrm{eff}} = \mathbf{H}_{0}+\mathbf{H}_{\mathrm{D}}+\mathbf{h}(t)+\ldots$ contains the external field $\mathbf{H}_{0}$, DMI field $\mathbf{H}_{\mathrm{D}} = H_{\mathrm{D}}\left[\mathbf{e}_{z}\times\mathbf{l}\right]$ where $\mathbf{d} = \mathbf{e}_{z}$ (in the case of the uniaxial magnet FeBO$_3$) 
and the THz magnetic field $\mathbf{h}(t).$ The first two terms in Eq.~\ref{eq:3} describe the canting of the sublattices as a result of $\mathbf{H}_{0}$ and $\mathbf{H}_{\mathrm{D}}$. The last term describes the dynamic magnetization induced by a dynamic $\mathbf{l}(t)$, as was also written in Eq.~(1) of the main article, which is the key to the nonlinear THz control via the non-equilibrium magnonic state. The Lagrangian of the sigma model $\mathcal{L}$ for a common antiferromagnet, with or without canting, reads \cite{doi:10.1063/1.4865565}:
\begin{equation}\label{eq:4}
\frac{1}{\hbar }\mathcal{L =  }\frac{1}{2\gamma H_{\mathrm{ex}}}\left(\frac{\partial\mathbf{l}}{\partial t}\right)^{2}-\frac{1}{H_{\mathrm{ex}}}\left(\mathbf{H}\cdot\left[\mathbf{l}\times\frac{\partial\mathbf{l}}{\partial t}\right]\right) -\overline{w}_{\mathrm{a}}\left(\mathbf{l}\right),
\end{equation}
where $\mathbf{H} = \mathbf{H}_{\mathbf{0}}+\mathbf{h}(t)$. The first term can be interpreted as classical kinetic energy and the second term is a gyroscopic term. The effect of the canting by a DMI field is contained in the effective anisotropy energy $\overline{w}_{\mathrm{a}}\left(\mathbf{l}\right) = w_{\mathrm{a}}\left(\mathbf{l}\right)+\gamma^{2}\left[\left(\mathbf{H}_{\mathrm{eff}}\cdot\mathbf{l}\right)^{2}-\mathbf{H}_{\mathrm{eff}}^{2}\right]/2H_{\mathrm{ex}}$ where $w_{\mathrm{a}}\left(\mathbf{l}\right)$ is the usual anisotropy energy (in units of frequency). For biaxial crystals, the latter is given by:
\begin{equation}\label{eq:5}
w_{\mathrm{a}}\left(\mathbf{l}\right) =\frac{1}{2}\omega_{\mathrm{a}}l_{z}^{2}+\frac{1}{2}\omega_{\mathrm{p}}l_{y}^{2},
\end{equation}
where $\omega_{\mathrm{a}} = \gamma H_{\mathrm{a}}$ with $H_{\mathrm{a}}$ the uniaxial anisotropy field. For uniaxial FeBO$_3$, $\omega_{\mathrm{p}}$ could be induced by, for example, external stress or magneto-elastic interactions, but in general, it can be put to zero $\omega_{\mathrm{p}} = 0$. In this case, the ground state of FeBO$_3$ will be as depicted in Fig.~$2$ of the main article with $\mathbf{l}\parallel\mathbf{e}_{x}$ and $\mathbf{m}\parallel\mathbf{e}_{y}$, considering that $\mathbf{H}_0 \parallel \mathbf{e}_{y}$ aligns the magnetization. Note that in this theory, we ignore the small out-of-plane external magnetic field $z$-component which is negligible as explained in Section~\ref{sec:2Dspectroscopy}, and which is present in the experiment because of the $12^{\circ}$ tilt of the sample that was set to ensure a projection of $\mathbf{m}(t)$ for both modes along the optical axis for the purpose of detection using magneto-optics (see Ref.~\cite{PhysRevLett.123.157202}). 

In linear approximation, only terms bilinear over all the magnon amplitudes $l_{y}, l_{z}$ and linear over the THz magnetic field $\mathbf{h}(t)$ are considered, which is usual practice when considering magnetization dynamics. However, the nonlinear coupling of the two modes only exists within quasi-linear approximation; it is caused by bilinear terms of the form $l_{y}l_{z}h(t)$. To this end, it is convenient to make the substitution $l_{x}\mapsto 1-(l_{y}^{2}+l_{z}^{2})/2$ and write the Lagrangian (in quasi-linear approximation) in terms of $l_{y}$ and $l_{z}$ only, including bilinear terms:
\begin{equation}\label{eq:6}
\mathcal{L} \approx \mathcal{L}_{y,0}+\mathcal{L}_{z,0}+\mathcal{L}_{\mathrm{coupl}},
\end{equation}
where the first two terms are the same as in linear approximation, and $\mathcal{L}_{\mathrm{coupl}}$ describes the non-linear coupling of the modes, driven by the THz field $h(t)$. These terms are:
\begin{equation}\label{eq:7}
\frac{\gamma H_{\mathrm{ex}}}{\hbar }\mathcal{L}_{y,0} =\frac{1}{2}\left(\frac{\partial l_{y}}{\partial t}\right)^{2}-\frac{1}{2}\omega_{\mathrm{FM}}^{2}l_{y}^{2}+l_{y}\left[\gamma^{2}\left(H_{0}+H_{\mathrm{D}}\right)h_{x}+\gamma\frac{\mathrm{d}h_{z}}{\mathrm{d}t}\right], \end{equation}
\begin{equation}\label{eq:8}
\frac{\gamma H_{\mathrm{ex}}}{\hbar }\mathcal{L}_{z,0} =\frac{1}{2}\left(\frac{\partial l_{z}}{\partial t}\right)^{2}-\frac{1}{2}\omega_{\mathrm{AF}}^{2}l_{z}^{2}-\gamma l_{z}\frac{\mathrm{d}h_{y}}{\mathrm{d}t}, \end{equation}
\begin{equation}\label{eq:9}
\frac{\gamma H_{\mathrm{ex}}}{\hbar }\mathcal{L}_{\mathrm{coupl}} = \gamma^{2}H_{0}h_{z}l_{y}l_{z}+\gamma h_{x}\left(l_{z}\frac{\mathrm{d}l_{y}}{\mathrm{d}t}-l_{y}\frac{\mathrm{d}l_{z}}{\mathrm{d}t}\right),
\end{equation}
where $\omega_{\mathrm{FM}}$ and $\omega_{\mathrm{AF}}$ are the angular frequencies of the q-FM and q-AF modes, which are given in the general biaxial case by:
\begin{equation}\label{eq:10}
\omega_{\mathrm{FM}}^{2} = \gamma^{2}H_{0}\left(H_{0}+H_{\mathrm{D}}\right)+\gamma \omega_{\mathrm{p}}H_{\mathrm{ex}},  \quad \quad \quad \quad \omega_{\mathrm{AF}}^{2} = \gamma \omega_{\mathrm{a}}H_{\mathrm{ex}}+\gamma^{2}H_{0}H_{\mathrm{D}}. 
\end{equation}

Again, for uniaxial FeBO$_3$ we can put $\omega_{\mathrm{p}} = 0$. Note that in our experimental geometry, the THz magnetic field is predominantly polarized within the $x-y$ plane (again ignoring the $12^\circ$ canting of the sample) and thus we can in principle put $h_{z} = 0$ such that the coupling only originates from the gyroscopic term. We then write for the THz field components $h_{x} = h\cos \alpha$ and $h_{y} = h\sin \alpha$, where $h(t) = \vert\mathbf{h}(t)\mathbf{\vert }$ determines the shape of the THz pulse and $\alpha$ is the angle of the THz magnetic field w.r.t. the $x$-axis. The equations of motion, derived from the Euler-Lagrange equations, then take the form:
\begin{equation}\label{eq:11}
\frac{d^{2}l_{y}}{\mathrm{d}t^{2}}+ \omega_{\mathrm{FM}}^{2}l_{y} = \left[\gamma^{2}\cos \alpha\left(H_{0}+H_{\mathrm{D}}\right)h+\gamma\frac{\mathrm{d}h_{z}}{\mathrm{d}t}\right]-2\gamma h\cos \alpha\frac{\mathrm{d}l_{z}}{\mathrm{d}t}-\gamma^{2}H_{0}h_{z}l_{z}, 
\end{equation}
\begin{equation}\label{eq:12}
\frac{d^{2}l_{z}}{\mathrm{d}t^{2}}+ \omega_{\mathrm{AF}}^{2}l_{z} =\left[-\gamma \sin \alpha \frac{\mathrm{d}h}{\mathrm{d}t}\right]+2\gamma h\cos \alpha\frac{\mathrm{d}l_{y}}{\mathrm{d}t}-\gamma^{2}H_{0}h_{z}l_{y}. 
\end{equation}
Here the square brackets are used to indicate the contribution of the linear excitations. By ignoring the small tilt of the sample $h_{z} = 0$, the equations simplify to:
\begin{equation}\label{eq:13}
\frac{d^{2}l_{y}}{\mathrm{d}t^{2}}+\omega_{\mathrm{FM}}^{2}l_{y} = \gamma^{2}\cos \alpha\left(H_{0}+H_{\mathrm{D}}\right)h-2\gamma h\cos \alpha\frac{\mathrm{d}l_{z}}{\mathrm{d}t}, 
\end{equation}
\begin{equation}\label{eq:14}
\frac{d^{2}l_{z}}{\mathrm{d}t^{2}}+ \omega_{\mathrm{AF}}^{2}l_{z} =  -\gamma \sin \alpha\frac{\mathrm{d}h}{\mathrm{d}t}+2\gamma h\cos \alpha\frac{\mathrm{d}l_{y}}{\mathrm{d}t}.
\end{equation}

An accurate function $h = h(t)$ that describes the THz waveform is given by the Gaussian derivative function:
\begin{equation}\label{eq:15}
h(t) = t_{0}\frac{d}{\mathrm{d}t}f(t), \quad \quad \quad  f(t) = h_{0}\sin \omega_{0}t\exp\left(-\frac{t^{2}}{ t_{0}^{2}}\right),
\end{equation}
where $t_{0}\approx 0.36$~ps, $\omega_{0} = 1/t_{0}$ and $h_{0}$ the peak magnetic field strength of the THz pulse in Tesla. For Fig.~$4$(b) of the main article, we numerically solved these equations of motion using the parameters $h_{0} = 110$ mT, $H_{0} = 70$~mT, $\alpha  = \pi /4$, $\frac{1}{2\pi }\omega_{\mathrm{FM}} = f_{\mathrm{FM}} = 0.026$~THz, $\frac{1}{2\pi }\omega_{\mathrm{AF}} = f_{\mathrm{AF}} = 0.473$~THz, $\frac{\gamma }{2\pi } = 28$~GHz/T and finally $H_{\mathrm{D}} = 6.19$~T \cite{RUDASHEVSKY1972959}. We plotted the $2$D~Fourier transform of the nonlinear part of the sum of the two dynamical components $S\left(t, \tau\right)\equiv l_{y}\left(t, \tau\right)+l_{z}\left(t, \tau\right)$ (see Eq.~(2) of the main article with $S$ instead of $\theta$).

The equations of motion can also be solved analytically. For the case of short optical pulses, $\int h(t)\mathrm{d}t\neq 0$, and the driving field $h(t)$ can be replaced by the delta-function $h(t)\mapsto \delta(t)\int h(t)\mathrm{d}t$ that gives inertial dynamics with nonzero $\left.\frac{\mathrm{d}l}{\mathrm{d}t}\right\vert_{t = 0_{+}}$  \cite{Kimel2009}. Alternatively, $\frac{\mathrm{d}h(t)}{\mathrm{d}t}$ can be replaced by the derivative of the delta function, giving initial conditions for $\left.l(t)\right\vert_{t = 0_{+}}$, after which the free solution can be used given these initial conditions \cite{Galkin2008}. For a THz pulse, however, $\int h(t)\mathrm{d}t = 0$ and similar tricks do not work. Therefore, the equations should be solved more accurately, starting by making the complex transformation $\xi  =\frac{\mathrm{d}l_{z}}{\mathrm{d}t}+i\omega_{\mathrm{AF}}l_{z}$ and solving the resulting first-order differential equation. Initially, $l_{y,z} = 0,\frac{\mathrm{d}l_{y,z}}{\mathrm{d}t} = 0$, and the formal solution to Eq.~\eqref{eq:14} can be written in complex form \cite{LandauLifshitz}:
\begin{equation}\label{eq:16}
l_{z}(t) =\frac{1}{\omega_{\mathrm{AF}}}\mathfrak{I[} \xi ] =\frac{1}{\omega_{\mathrm{AF}}}\mathfrak{I}[-\gamma \sin \alpha \exp\left(i\omega_{\mathrm{AF}}t\right)\int_{-\infty }^{t}\frac{\mathrm{d}h\left(t'\right)}{\mathrm{d}t'}e^{-i\omega_{\mathrm{AF}}t'}\mathrm{d}t']. 
\end{equation}
At large times $t\gg t_{0}$, the upper limit in the integral can be replaced by $t = \infty$ and the solution can be obtained by repeated integration by parts:
\begin{equation}\label{eq:17}
\left.l_{z}(t)\right\vert_{t\gg t_{0}} = -\gamma \sin \alpha h_{0}\omega_{\mathrm{AF}}t_{0}^{2}\Xi\left(\omega_{\mathrm{AF}}\right)\cos \omega_{\mathrm{AF}}t, \\ 
\end{equation}
where $\Xi(\omega) = \sqrt{\pi} e^{-\frac{1}{4}(1+t_0^2\omega^2)} \sinh(\frac{1}{2}t_0\omega)$. These oscillations have a considerable amplitude given the frequency of the q-AF mode $\frac{1}{2\pi }\omega_{\mathrm{AF}}\approx 0.5$~THz. On the contrary, a similar linear excitation via linear torque on $l_{y}(t)$ describing the q-FM resonance mode gives:
\begin{equation}\label{eq:18}
\left.l_{y}(t)\right\vert_{t\gg t_{0}} = \gamma^{2}\cos \alpha\left(H_{0}+H_{\mathrm{D}}\right)h_{0}t_{0}^{2}\Xi (\omega_{\mathrm{FM}})\sin \omega_{\mathrm{FM}}t.
\end{equation}
For the q-FM mode with $\frac{1}{2\pi }\omega_{\mathrm{FM}}\sim 0.03$ THz, the maximum amplitude is approximately 40 times weaker compared to that of the q-AF mode described by Eq.~\eqref{eq:17}. Instead, for an efficient excitation of the q-FM mode we should consider in Eq.~\eqref{eq:13} the second term proportional to $\frac{\mathrm{d}l_{z}}{\mathrm{d}t}$, which captures the enhanced susceptibility of this mode to the THz magnetic field through the q-AF mode. An expression for this magnon-mediated driving force by the second THz pulse, with field $h_{0}$, polarization angle $\alpha'$ and delayed at a time $\tau$ w.r.t. the first pulse, is found by substituting Eq.~\eqref{eq:17} into \eqref{eq:13} and solving the resulting equation of motion. The solution is given by:
\begin{equation}\label{eq:19}
\begin{split}
\left.l_{y}\left(t, \tau\right)\right\vert_{t\gg \tau } =  \gamma^{2}\sin \alpha \cos \alpha 'h_{0}^{2}t_{0}^{4}\omega_{\mathrm{AF}}^{2}\omega_{\mathrm{FM}}^{-1}\Xi\left(\omega_{\mathrm{AF}}\right) \\ 
\cdot[\left(\omega_{+}\Xi\left(\omega_{+}\right)\sin \omega_{+}\tau +\omega_{-}\Xi\left(\omega_{-}\right)\sin \omega_{-}\tau\right)\sin \omega_{\mathrm{FM}}t \\ +\left(\omega_{+}\Xi\left(\omega_{+}\right)\cos \omega_{+}\tau -\omega_{-}\Xi\left(\omega_{-}\right)\cos \omega_{-}\tau\right)\cos \omega_{\mathrm{FM}}t] 
\end{split}
\end{equation}
Where $\omega_{\pm } = \omega_{\mathrm{AF}}\pm \omega_{\mathrm{FM}}$. When $\omega_{\mathrm{FM}}\ll \omega_{\mathrm{AF}}$, we have that $\omega_{+} \approx \omega_{-}$ such that the second term $\sim \cos \omega_{\mathrm{FM}}t$ can be ignored. As a side remark, note that the periodicity of excitation occurs with the excitation frequencies $\omega_{\pm } = \omega_{\mathrm{AF}}\pm \omega_{\mathrm{FM}}$ which is Doppler-shifted, just as we observed experimentally and which is explained intuitively in Section~\ref{sec:geometry}. Now, we can give a quantitative expression of how much a dynamic magnetization $l_{z}(t)$ induced by a first ``preparation pulse'' polarized along the $y$-axis ($\alpha  = \pi /2)$ can enhance the susceptibility of a second ``excitation pulse'' with polarization along the $x$-axis ($\alpha' = 0$). Without the dynamic magnetization present, the excitation of the second pulse would just be the linear one from Eq.~\eqref{eq:18}, but the presence of the dynamic magnetization gives an additional contribution as described by Eq.~\eqref{eq:19}. The amplification factor $\mu$ when $\omega_{\mathrm{FM}}\ll \omega_{\mathrm{AF}}$ can then be expressed:
\begin{equation}\label{eq:40}
\mu \approx 1+\frac{h_{0}t_{0}^{2}\omega_{\mathrm{AF}}^{2}\Xi\left(\omega_{\mathrm{AF}}\right)\left(\omega_{+}\Xi\left(\omega_{+}\right)+ \omega_{-}\Xi\left(\omega_{-}\right)\right)}{\omega_{\mathrm{FM}}\left(H_{0}+H_{\mathrm{D}}\right)\Xi\left(\omega_{\mathrm{FM}}\right)} 
\end{equation}
The analytical expressions for the amplitudes and the amplification factor are matching very well to the values obtained when solving the differential equations~\eqref{eq:13} and \eqref{eq:14} numerically. In our case, with THz magnetic field $\sim 0.1$~T, the amplification factor is about $\mu \sim 7$. But for systems with low DMI, this amplification could become several orders of magnitude. Note that the resulting expression for the amplification factor could depend on the THz pulse shape (see Eq.~\eqref{eq:15}), in which case the analysis should be repeated to obtain the correct amplification factor. However, the general mechanism is completely universal. 

Alternatively, also a single THz pulse (as in Ref. \cite{PhysRevLett.123.157202}) is known to be capable of exciting spin-dynamics through this magnon-mediated mechanism, but then we need to consider the solution within the pulse action, i.e., at times $t\leq t_{0}$. Using Eq.~\eqref{eq:14}, the excitation of the q-AF mode is approximately described by: 
\begin{equation}\label{eq:41}
\left.\frac{\mathrm{d}l_{z}}{\mathrm{d}t}\right\vert_{t\leq t_{0}}\approx h(t)\gamma \sin \alpha,
\end{equation}
which can be substituted in Eq.~\eqref{eq:13} to act as a driving force for the q-FM mode. Although it is not possible to obtain the exact solution, the amplitude will again be quadratic with the THz field $\sim h^{2}$ and proportional to $\sin \alpha \cos \alpha$, as was observed in the experiments. It should be noted that this model and the mechanism of excitation can be applied to any antiferromagnet, with or without canting, external magnetic field, or any other factors that lower the dynamic symmetry. Therefore, although we focus on the case of FeBO$_3$, we arrived at a universal mechanism of excitation applicable to any antiferromagnet.
\newpage

\section{Supplemental data and figures}
\subsection{Classical illustration of experiment}
\begin{figure}[h!]
    \centering
    \includegraphics{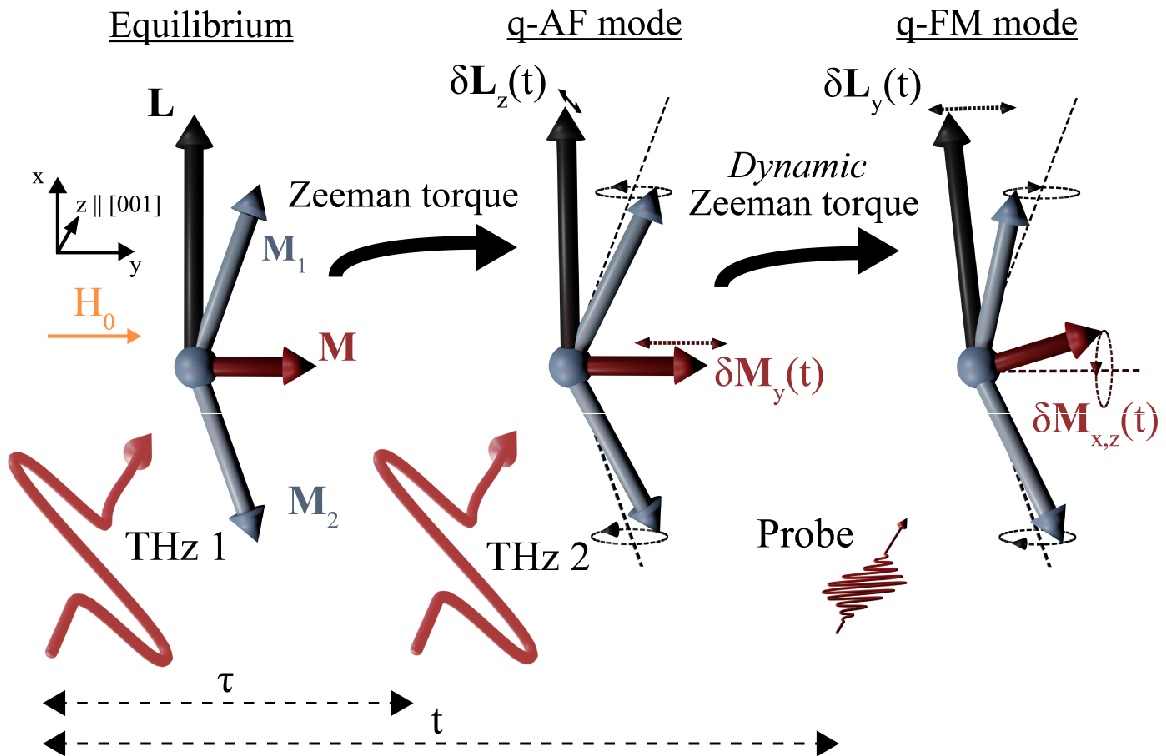}
    \caption{\textit{\small{General idea of the experiment. By applying a small external magnetic field of $70$~mT in the ($001$) easy plane we align the net magnetization. Both the applied THz magnetic field pulses are polarized at $45^\circ$. Ideally, the first polarization would be along $\mathbf{M}$ while the other one is perpendicular to it, however, the experimental setup only allows us to grant the same polarization to both pulses. The THz magnetic field component of the first THz pulse that is perpendicular to $\mathbf{L}$ will efficiently excite the q-AF mode ($f_{\mathrm{AF}} \approx 473$~GHz) by Zeeman torque, causing longitudinal modulations of the net magnetization $\delta M_y(t)$. Subsequently, the magnetic field component of the second THz pulse that is perpendicular to $\mathbf{M}$ can now couple efficiently to this emergent dynamic magnetization $\delta M_y(t)$ and thereby excite the q-FM mode ($f_{\mathrm{FM}} \approx 26$~GHz). This can be thought of as a dynamic analog of the THz Zeeman torque, as the THz magnetic field now couples to a dynamic magnetization instead of a static magnetization. The dynamics of both modes are detected by tracking the time-resolved polarization rotation of an optical probe pulse as a function of the delay time $t$.}}}
    \label{fig:ideaexperiment}
\end{figure}
\newpage 
\subsection{Supplemental figures of the experiment and simulation}
In the main article, we have only considered the total signal $\theta_{12}(t,\tau)$ (Fig.~$3$, main article) and the $2$D~FFT of the non-linear part $\tilde{\theta}_{\mathrm{NL}}(f_{\mathrm{det}}, f_{\mathrm{ex}})$ (Fig.~$4$(a), main article). Regarding the numerical simulations, we only plotted the 2D~FFT of the nonlinear part $\tilde{S}_{\mathrm{NL}}(f_{\mathrm{det}},f_{\mathrm{ex}})$ in Fig.~$4$(b) of the main article. Here, we also show time-resolved simulations for both $S_{12}(t,\tau)$ and $S_{\mathrm{NL}}(t,\tau)$ as well as the $2$D~FFT $\tilde{S}_{12}(f_{\mathrm{det}},f_{\mathrm{ex}})$. In summary, in this Section, we plot all possible figures associated with the experiment and simulation, both in frequency and time domain. 

\begin{figure}[h!]
    \centering
    \includegraphics{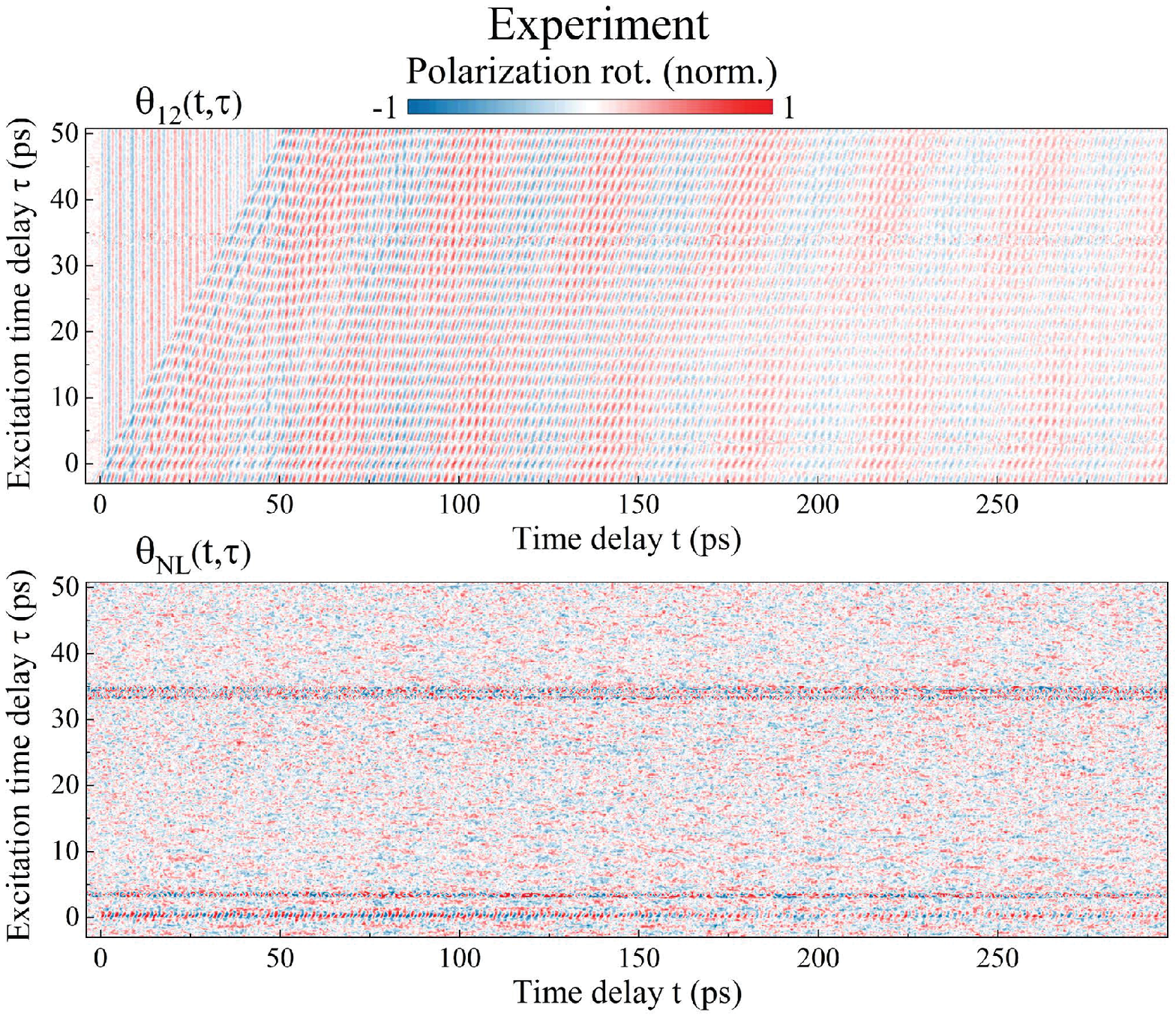}
    \caption{\textit{\small{Total and nonlinear signal of experiments in time-domain. (Upper panel) The double-pump induced polarization rotation when both THz pulses are present. (Lower panel) Extracted nonlinear component of the signal. We attribute the dark lines at the excitation time delays at about $4$~ps and between $30$ and $40$~ps to instabilities of the experimental setup during the measurements, the measurements which undertook nearly $48$ hours, but these artifacts do not influence the further analysis of the data and subsequent results and conclusions. The dark line at $\tau = 0$ is where the two pulses are simultaneously generated inside the LiNBO$_3$ crystal, here superposition of the THz pulses itself no longer holds, and we exclude this part from the analysis (as explained in Section~\ref{sec:2Dspectroscopy}).  }}}
    \label{fig:Exp_TD}
\end{figure}

\newpage

\begin{figure}[h!]
    \centering
    \includegraphics[width = 0.9\textwidth]{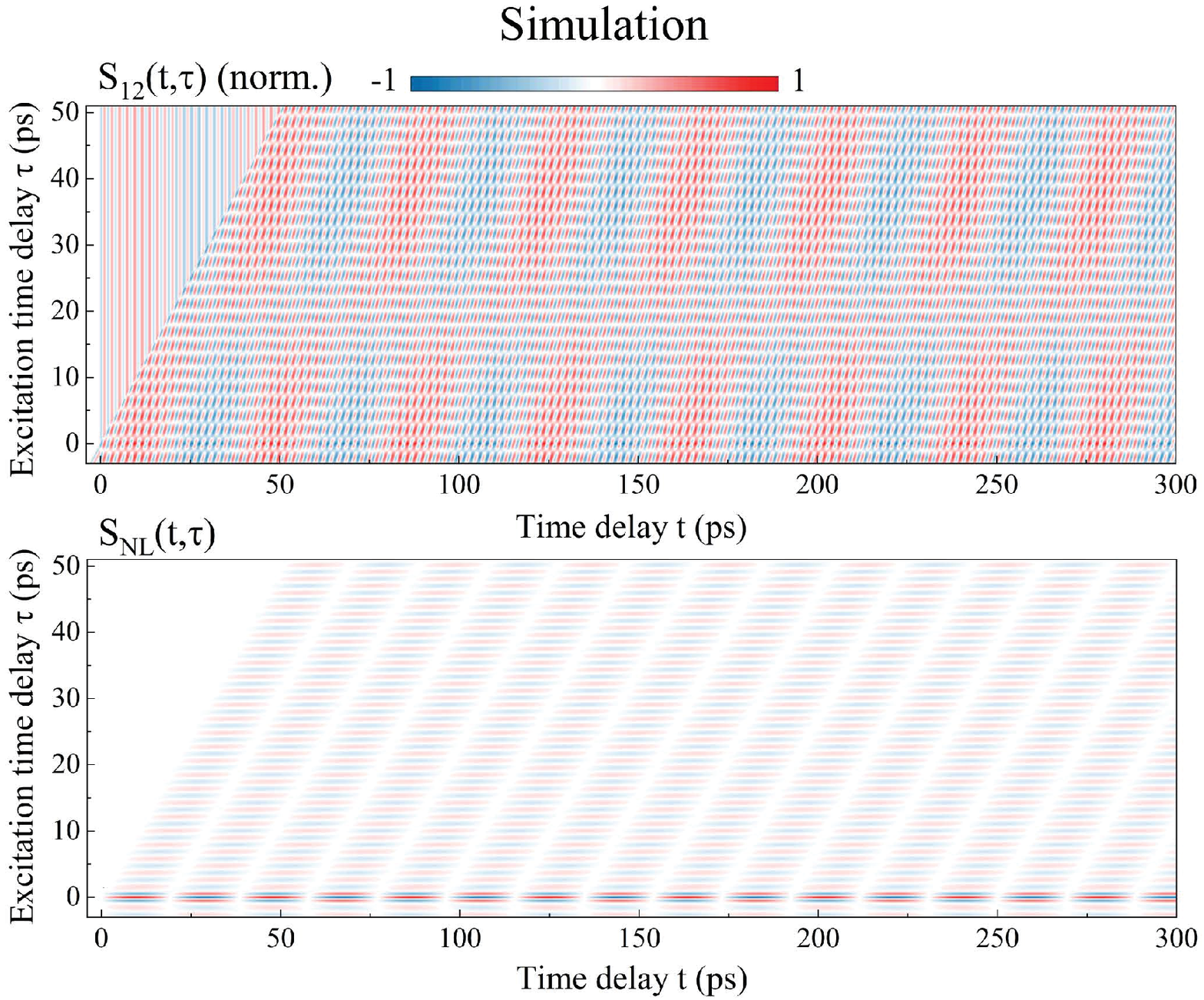}
    \caption{\textit{\small{Total and nonlinear signal of the simulated data in time-domain. (upper panel) Total signal of the simulated data $S(t,\tau) \equiv l_y(t,\tau) + l_z(t,\tau)$, where both the simulated THz pulses of magnetic field are present. (lower panel) Nonlinear part of the simulated signal, which is non-zero due to the coupling term in the quasi-linear Lagrangian (Eq.~\eqref{eq:6}). }}}
    \label{fig:Sim_TD}
\end{figure}

\begin{figure}[h!]
    \centering
    \includegraphics{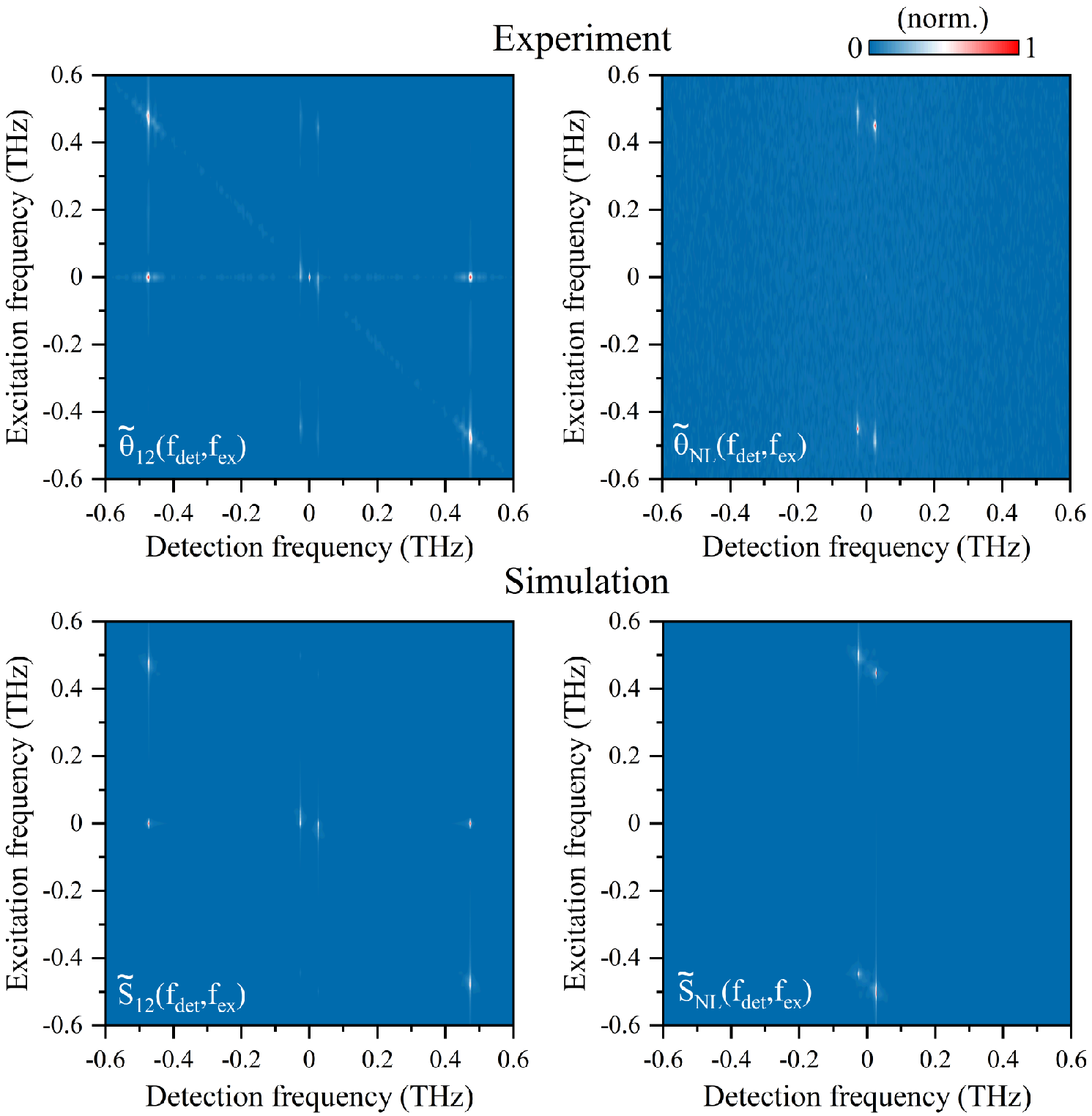}
    \caption{\textit{\small{$2$D~FFTs of the experimental and simulated data. For all subfigures, we are only required to examine the first (I) and fourth (IV) quadrants as the data which is transformed is purely real which implies that the Fourier amplitude has inversion symmetry $|\tilde{\theta}_{\mathrm{NL}}(f_{\mathrm{det}}, f_{\mathrm{ex}})| = |\tilde{\theta}_{\mathrm{NL}}(-f_{\mathrm{det}}, -f_{\mathrm{ex}})|$. (upper left) The horizontal line at $f_{ex} = 0$ shows by definition the single-pulse spectrum, and contains two pronounced peaks at the detection frequency corresponding to the q-FM and q-AF modes. The small dot in the center represents DC contributions to the signal. Furthermore, the diagonal line in the fourth quadrant represents all the linear contributions, terminating at a peak at a frequency of $(f_{\mathrm{det}}, f_{\mathrm{ex}}) = (f_{\mathrm{AF}}, - f_{\mathrm{AF}})$ which reflects the linear Zeeman excitation of the q-AF mode by the THz field component $\tilde{h}(\omega_{\mathrm{AF}})$. The faint horizontal and diagonal lines are a result of the overlap of the THz pulse with the probe pulse and can be interpreted as the Pockels effect where the direct presence of the THz electric field changes the birefringence of the medium linearly. This is why in the (bottom left) panel, where we plot the $2$D~FFT of the total simulated signal, these lines are not present, as we do not account for the Pockels effect in the simulations. Here, we only see the peaks associated with the q-FM and q-AF resonance modes, and the off-diagonal peaks corresponding to the coupling of these modes. The $2$D~FFTs of the non-linear parts of the experimental and simulated data (upper and bottom right), which are also shown in the main article, clearly demonstrate the non-linear channel of energy transfer from the q-AF to the q-FM mode. The large degree of similarity between the experiment and theory demonstrates that this nonlinear effect is very well described by the Lagrangian theory. }}}
    \label{fig:suppl_FD}
\end{figure}

\clearpage

\bibliography{references}